\documentclass[conference, a4paper]{IEEEtran}
\usepackage[nolist]{acronym}
\usepackage{caption}
\usepackage{subfig}
\usepackage{graphicx}
\usepackage{tabularx}
\usepackage{amsmath,amssymb}
\usepackage{cite}
\usepackage[usenames,dvipsnames]{pstricks}
\usepackage{epsfig}
\usepackage{pst-grad} 
\usepackage{pst-plot} 
\usepackage[utf8]{inputenc}
\usepackage{epsfig,psfrag}
\usepackage{subfig}
\usepackage{bm}
\usepackage{xcolor,import}
\ifCLASSINFOpdf
\else
\fi
\author{\IEEEauthorblockN{Quentin Bodinier,
		Faouzi Bader, and
		Jacques Palicot} 
	\IEEEauthorblockA{SCEE/IETR - CentraleSupélec, Rennes, France, \\}
	\IEEEauthorblockA{Email : \{firstname.lastname\}@supelec.fr\\}}
\begin{document}
	\DeclareGraphicsExtensions{eps}
	\graphicspath{{fig/}}
	\title{Modeling Interference Between OFDM/OQAM and CP-OFDM: Limitations of the PSD-Based Model}
	
	\maketitle
	
	\begin{abstract}
		To answer the challenges put out by the next generation of wireless networks (5G), important research efforts have been undertaken during the last few years to find new waveforms that are better spectrally localized and less sensitive to asynchronism effects than the widely deployed Cyclic Prefix Orthogonal Frequency Division Multiplexing (CP-OFDM). One of the most studied schemes is OFDM-Offset Quadrature Amplitude Modulation (OFDM/OQAM) based on the PHYDYAS filter pulse. In the recent literature, spectrum coexistence between OFDM/OQAM and CP-OFDM is commonly studied based on the Power Spectral Density (PSD) model. 
		In this paper, we show that this approach is flawed and we show that the actual interference injected by OFDM/OQAM systems onto CP-OFDM is much higher than what is classically expected with the PSD based model in the literature. We show that though using OFDM/OQAM in secondary systems is still advantageous, it brings limited gain in the context of coexistence with incumbent CP-OFDM systems. 
	\end{abstract}
	

	\IEEEpeerreviewmaketitle

	\section{Introduction}
	\label{sec:intro}
	The advent of the 5th Generation of wireless communication systems (5G) is envisioned to bring flexibility to cellular networks. New services as Device-To-Device (D2D) or Machine-To-Machine (M2M) communications are expected to be massively deployed in the near future. Such new communication devices have to coexist with incumbent legacy systems in the cell, i.e. Long-Term-Evolution Advanced (LTE-A) users. In such heterogeneous environments, perfect synchronization between the different types of systems is not feasible. This loss of synchronization will cause harmful interference between active users, which will in turn degrade the overall system performance. 

This hurdle can be overcome through the design of new waveforms that are robust against asynchronism, and well localized in both time and frequency domains. As a matter of fact, it is now widely accepted that the Cyclic Prefix-Orthogonal Frequency Division Multiplexing (CP-OFDM) used in LTE-A is not adapted for flexible sharing and coexistence in fragmented spectrum for heterogeneous networks \cite{Lahetkangas2013,wunder2014}. Indeed, as soon
as the orthogonality between CP-OFDM users is destroyed, for example because of the coexistence between unsynchronized incumbent and secondary systems, their performance shrinks dramatically \cite{Medjahdi2011}. This is mainly due to the fact that CP-OFDM systems filter symbols with a time-rectangular window, which causes poor frequency localization \cite{Farhang-Boroujeny2011, Baltar2007,Mahmoud2009} and high asynchronism sensitivity in the multi-user context \cite{Raghunath2009,Mahmoud2009, Aminjavaheri2015, Speth1999}.

OFDM with Offset Quadrature Amplitude Modulation (OFDM/OQAM) \cite{Bellanger2010}, is one of the main new waveform schemes explored by the research community. Indeed, it overcomes the cited CP-OFDM limitations and enables both higher flexibility and reduction of interference leakage for multi-standard systems coexistence \cite{Farhang-Boroujeny2011, Baltar2007,Mahmoud2009}. The coexistence between OFDM/OQAM based D2D pairs and CP-OFDM LTE users has been widely studied in \cite{Xing2014, BodinierICC2016}.

To the best of the authors knowledge, in all studies on coexistence between OFDM/OQAM secondary users \cite{Xu2008, shaat2010computationally, skrzypczak2012ofdm} and CP-OFDM incumbent ones, the interference caused by the different types of users onto each other is quantified with the Power Spectral Density (PSD)-based model originally proposed in \cite{weiss04}. Yet, the authors pointed out in \cite{BodinierICC2016} that values of interference obtained by means of Monte-Carlo simulations were much higher than those obtained with the PSD-based model.
In \cite{Medjahdi2010a}, Medjahdi \textit{et. al.} designed a more precise interference model named "instantaneous interference" that takes into account the demodulation operations and the time asynchronism between users. Nevertheless, the aforementioned study only analyzed the multiuser interference in cases where all users are using the same waveform, either CP-OFDM or OFDM/OQAM. No such analysis has been applied to heterogeneous scenarios where CP-OFDM and OFDM/OQAM system are deployed in the same geographical area and coexist in the same cell.

The approach of this paper is therefore to study inter-user interference in scenarios where CP-OFDM and OFDM/OQAM users interfere with each other. It is shown that the PSD based approach consists in modeling the interference at the input antenna of the interfered receiver, and totally omits the demodulation operations that are performed by the latter. We show that the actual interference seen at the output of the demodulator of the interfered receiver is much higher than expected using the PSD based model. Moreover, we show that interference between the incumbent and secondary systems is symmetrical, which contradicts the results obtained with the PSD-based model. Finally, the presented study nuances results classically shown in the literature, and diminishes the benefits expected from using OFDM/OQAM for coexistence with CP-OFDM incumbent systems.

The remainder of this paper is organized as follows: Section \ref{sec:model} presents the system model. In Section \ref{sec:bcg}, a short overview on CP-OFDM and OFDM/OQAM systems is given. In Section \ref{sec:interf}, the different models used to rate heterogeneous interference between OFDM/OQAM and CP-OFDM are presented. In Section \ref{sec:results}, numerical results are presented and 
concluding remarks are provided in Section \ref{sec:ccl}.\\

\textit{Notations:} Throughout this paper, scalars are noted $x$, vectors are bold-faced as $\mathbf{x}$, $k$ represents the discrete time sample index, $n$ indexes symbols and $m$ indexes subcarriers. $\mathcal{R}\{.\}$ is the real part operator and $E_\alpha\{.\}$ is the mathematical expectation with respect to the random variable $\alpha$.
	
	\section{System Model}
	\label{sec:model}
	\begin{figure}
	\includegraphics[width=\linewidth]{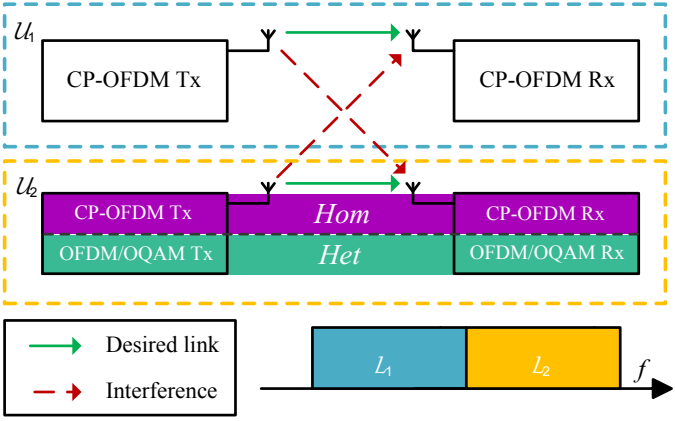}
	\caption{Summary of the study led in the paper : two users $\mathcal{U}_1$ and $\mathcal{U}_2$ transmit on adjacent bands $\mathcal{L}_1$ and $\mathcal{L}_2$ and interfere with each other. Channel is assumed perfect and no Gaussian noise is considered. $\mathcal{U}_2$ uses CP-OFDM in \textit{Hom} scenario and OFDM/OQAM in \textit{Het} scenario.}
	\label{fig:scenario_all}
	\vspace{-15pt}
\end{figure}

In this paper, we focus on rating the inter-user interference caused by the fact that different users transmit in an asynchronous manner with different waveforms. Therefore, we consider a simple scenario where an incumbent system  $\mathcal{U}_1$ coexists with a secondary user $\mathcal{U}_2$ in the same band. Both systems use multicarrier waveforms with the same subcarrier spacing $\Delta F$, and each of them is assigned a set of subcarriers $\mathcal{L}_i$.  The incumbent $\mathcal{U}_1$ utilizes CP-OFDM, whereas two alternatives are studied for $\mathcal{U}_2$. The latter uses CP-OFDM in the case of a homogeneous scenario (referred to as \textit{Hom}) and OFDM/OQAM in the case of a heterogeneous scenario (referred to as \textit{Het}). The configurations studied in this paper are summarized in Fig.~\ref{fig:scenario_all}. To focus the study on interference coming from the coexistence between these two systems, all channels are assumed perfect, and no Gaussian noise is considered.
Considering an infinite transmission on $M$ subcarriers, the sequences of symbols estimated at the receiver of $\mathcal{U}_1$ and $\mathcal{U}_2$ are modeled by 
\begin{eqnarray}
\hat{\mathbf{d}}_{1,m}[n] &=& \mathbf{d}_{1,m}[n]+\boldsymbol{\eta}_m^{2\rightarrow 1}[n],\label{eq1}\\
\hat{\mathbf{d}}_{2,m}[n] &=& \mathbf{d}_{2,m}[n]+\boldsymbol{\eta}_m^{1\rightarrow 2}[n],\label{eq2}\\\nonumber&&
\forall n\in\mathbb{N},\ \forall m=0\ldots M-1
\end{eqnarray}
where $\mathbf{d}_{i,m}[n]$ is the $n$-th symbol transmitted on the $m$-th subcarrier by user $\mathcal{U}_i$, and $\boldsymbol{\eta}_m^{j\rightarrow i}[n]$ represents the interference injected by the user $\mathcal{U}_j$  onto the $n$-th time slot and $m$-th subcarrier of user $\mathcal{U}_i$.

\begin{figure}
	\vspace{-10pt}
	\includegraphics[width=\linewidth]{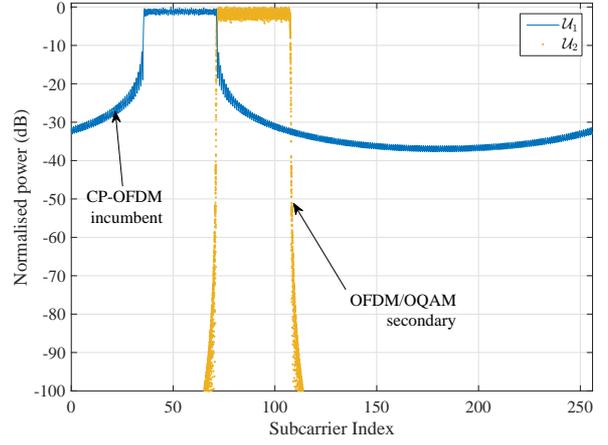}
	\caption{Spectral representation of the \textit{Het} scenario. CP-OFDM $\mathcal{U}_1$ and OFDM/OQAM $\mathcal{U}_2$ use directly adjacent bands $\mathcal{L}_1$ and $\mathcal{L}_2$ with same number of subcarriers.}
	\label{fig:scenario}
\end{figure}
In both analyzed scenarios, incumbent $\mathcal{U}_1$ and secondary $\mathcal{U}_2$ experience a loss of synchronization in time
domain. Besides, the time duration between two subsequent CP-OFDM symbols transmitted by the incumbent system $\mathcal{U}_1$ is $T_\text{s}+T_\text{CP}$, where $T_\text{s}$ is the time-symbol and $T_\text{CP}$ accounts for the duration of the CP. It is assumed that the transmission of $\mathcal{U}_2$ starts with a delay $\tau$ with respect to the transmission of $\mathcal{U}_1$. $\tau$ is taken as a random variable uniformly distributed in the interval $ [-\frac{T_\text{s}+T_\text{CP}}{2} \ \frac{T_\text{s}+T_\text{CP}}{2}[$. Therefore, the interference injected by the users onto each other is a function of the symbols they transmit and of the value of $\tau$. The mean interference power seen by each user on their $m$-th subcarrier is expressed as
\begin{eqnarray}
\small
I_m^{1\rightarrow 2} &=& E_{\mathbf{d}_1,\tau}\{|\boldsymbol{\eta}_m^{1\rightarrow 2}[n]|^2\}, \label{eq:i12}\\
I_m^{2\rightarrow 1} &=& E_{\mathbf{d}_2,\tau}\{|\boldsymbol{\eta}_m^{2\rightarrow 1}[n]|^2\}, 
\label{eq:i21}
\end{eqnarray}
and the total interference injected by each user onto the other is
\begin{eqnarray}
I^{1\rightarrow 2} &=& \sum_{m \in \mathcal{L}_2} I_m^{1\rightarrow 2}\label{eq:I12}\\
I^{2\rightarrow 1} &=& \sum_{m \in \mathcal{L}_1} I_m^{2\rightarrow 1}\label{eq:I21}
\end{eqnarray}
In the following, the structures of the CP-OFDM and OFDM/OQAM signals are briefly presented.

	\section{CP-OFDM and OFDM/OQAM PHY Characteristics}
	\label{sec:bcg}
	\subsection{CP-OFDM}
We consider a CP-OFDM system composed of $M$ subcarriers out of which $M_a$ are active. We define $M$ vectors $\mathbf{d}_m$ such that $\mathbf{d}_m$ is constituted of complex Quadrature Amplitude Modulation (QAM) symbols if subcarrier $m$ is active. Else,  $\mathbf{d}_m[n] = 0, \forall n \in \mathbb{N}$.
$N_\text{CP}$ being the length of the CP, the $n$-th OFDM symbol is expressed as

{\small\begin{eqnarray}
\mathbf{x}_n[k] &=& \sum_{m=0}^{M-1}\mathbf{d}_m[n]e^{j2\pi\frac{m}{M}k},\label{eq:OFDM_sig}\\&&n(M+N_{CP})-N_{CP} \leq k \leq n(M+N_{CP})+M-1,\nonumber
\end{eqnarray}}
and the total signal is expressed as $\mathbf{x}[k] = \sum_n \mathbf{x}_n[k]$.

To highlight the effects of inter-user interference only, we consider that the channel is perfect and that the CP-OFDM signal is polluted by an additive interfering signal $\mathbf{y}$. In that case, the $n$-th CP-OFDM estimated symbol is
\begin{eqnarray}
\hat{\mathbf{d}}_m[n] &=& \mathbf{d}_m[n] + \underset{\boldsymbol{\eta}_m[n]}{\underbrace{\sum_{k=n(M+N_{CP})}^{n(M+N_{CP})+M-1}\mathbf{y}[k]e^{j2\pi\frac{k}{M}m}}},\label{eq:OFDM_demod} \\&& \nonumber 0\leq m \leq M-1
\end{eqnarray}
where $\boldsymbol{\eta}_m[n]$ represent the total amount of interference that affects the estimated signal $\hat{\mathbf{d}}_m[n]$ as defined in (\ref{eq1}) or (\ref{eq2}).
\subsection{OFDM/OQAM}
The OFDM/OQAM system is composed of $M$ subcarriers out of which $M_a$ are active, where  $M$ vectors $\mathbf{d}_m$ contain real Pulse Amplitude Modulation (PAM) symbols if subcarrier $m$ is active and $\mathbf{d}_m[n] = 0, \forall n \in \mathbb{N}$ otherwise. A phase factor $\bm{\theta}_n[m] = e^{j\frac{\pi}{2}\lfloor\frac{n+m}{2}\rfloor}$ is further added to the symbols $\mathbf{d}_m[n]$. OFDM/OQAM is a based on a uniform polyphase filter bank structure 
with a prototype filter $\mathbf{g}$ of length $L_\mathbf{g} = KM$, where $K$ is called the overlapping factor, which is shifted to cover the whole of the system bandwidth. Subsequent symbols are separated by $\frac{M}{2}$ samples and are filtered through time-frequency shifted versions of $\mathbf{g}$. 
Therefore, each subcarrier is filtered by a filter $\mathbf{f}_m$ defined as : 
\begin{equation}
\mathbf{f}_m[k] = \mathbf{g}[k]e^{j2\pi\frac{m}{M}(k-\frac{KM-1}{2})}, 0 \leq k \leq KM-1
\end{equation}
and the $n$-th modulated OFDM/OQAM symbol is written as 
\begin{eqnarray}
\mathbf{x}_n[k] &=& \sum_{m=0}^{M-1} \mathbf{d}_n[m]\bm{\theta}_n[m]\mathbf{g}[k-n\frac{M}{2}]\times e^{j2\pi\frac{m}{M}(k-\frac{KM-1}{2})}, \nonumber\\
&& (n-K)\frac{M}{2} \leq k \leq (n+K)\frac{M}{2}-1\label{eq:FBMC_sig}
\end{eqnarray}

In this study, $\mathbf{g}$ is taken as the PHYDYAS filter with overlapping factor $K = 4$ \cite{Bellanger}. The frequency response of $g$ is expressed as
\begin{equation}
G(f) = \sum_{k=-(K-1)}^{K-1} G_{|k|}\frac{\sin(\pi(f-\frac{k}{KM})KM)}{KM \pi(f-\frac{k}{KM})}, 
\end{equation}
where $G_{0}=1$, $G_{1}=0.971960$, $G_{2}=1/\sqrt{2}$, and $G_{3} = 0.235147$ (see \cite{Bellanger} for more details on OFDM/OQAM modulation).

At the receiver, each subcarrier is filtered through the matched filter $\tilde{\mathbf{f}}_m$ and the real part of the signal is taken to remove purely imaginary intrinsic interference \cite{Farhang-Boroujeny2011} generated by the prototype filter $\mathbf{g}$. Therefore, the estimated signal
at the $m$-th subcarrier of the $n$-th symbol can be expressed as (\ref{eq:FBMCmod}).

\begin{figure*}[t]
	\begin{eqnarray}
	\hat{\mathbf{d}}_n[m] &=& \mathbf{d}_n[m]+\underset{\boldsymbol{\eta}_m[n]}{\underbrace{\mathcal{R}\left\{\sum\limits_{k=(n-K)\frac{M}{2}}^{(n+K)\frac{M}{2}}\mathbf{y}[k]\sum\limits_{\nu=-2K+1}^{2K-1}(-1)^{m(\nu-n)}\times e^{-j2\pi \frac{m}{M}(k-\frac{KM-1}{2})}\mathbf{g}\left[k+(\nu-n)\frac{M}{2}\right]\vphantom{\sum_{k=(n-K)\frac{M}{2}}^{(n+K)\frac{M}{2}}}\right\}}}\\\hline\nonumber
	\label{eq:FBMCmod}
	\end{eqnarray}
\end{figure*}

Based on the signal models defined in this section, we will hereafter discuss the modeling of the interference that the primary and secondary systems $\mathcal{U}_1$ and $\mathcal{U}_2$ inject onto each other.
	
	\section{Modeling Heterogeneous Interference}
	\label{sec:interf}
	\subsection{Mean Interference}
The amount of interference suffered by $\mathcal{U}_1$ and $\mathcal{U}_2$ on each of their subcarriers can be estimated from (\ref{eq:i12}),(\ref{eq:i21}). In the \textit{Hom} scenario, both  $\boldsymbol{\eta}^{1\rightarrow 2}$ and $\boldsymbol{\eta}^{2\rightarrow 1}$ are obtained by replacing $\mathbf{y}$ in (\ref{eq:OFDM_demod}) with $\mathbf{x}_n[k]$ expression of (\ref{eq:OFDM_sig}). Then, $I^{1\rightarrow 2}$ and $I^{2\rightarrow 1}$ are obtained by substituting (\ref{eq:OFDM_demod}) in (\ref{eq:I12}) and (\ref{eq:I21}) respectively. These derivations lead to the following expressions of the interference caused by $\mathcal{U}_1$ (resp. $\mathcal{U}_2$) onto $\mathcal{U}_2$ (resp. $\mathcal{U}_1$):
\begin{eqnarray}
I^{1\rightarrow 2}_{\textit{Hom}} &=& \sigma_{\mathbf{d}_1^2} \sum_{m \in \mathcal{L}_2, q \in \mathcal{L}_1} I^{1\rightarrow 2}_{\textit{Hom}}(q-m),\\
I^{2\rightarrow 1}_{\textit{Hom}} &=& \sigma_{\mathbf{d}_2^2} \sum_{m \in \mathcal{L}_1, q \in \mathcal{L}_2} I^{2\rightarrow 1}_{\textit{Hom}}(q-m),
\end{eqnarray}
where $\sigma_{\mathbf{d}_i^2}$ is the variance of $\mathbf{d}_i$. Besides, $\forall l$, $I_{\textit{Hom}}^{1\rightarrow 2}(l)$ (resp. $I_{\textit{Hom}}^{2\rightarrow 1}(l)$) represents the interference injected by the signal on the $q$-th subcarrier of $\mathcal{U}_1$ (resp. $\mathcal{U}_2$) onto the $m$-th subcarrier $m$ of (resp. $\mathcal{U}_1$) where $l = q-m$ is called the spectral distance. In \textit{Hom} scenario, $\forall l, I_{\textit{Hom}}^{1\rightarrow 2}(l) = I_{\textit{Hom}}^{2\rightarrow 1}(l)$, and Mejdahdi \textit{et. al.} have derived in \cite{Medjahdi2010a} a closed-form of the interference $I_{\textit{Hom}}^{1\rightarrow 2}(l)$ and tabulated its values in so-called "Mean Interference Tables".

In \textit{Het} scenario, the expression of $\boldsymbol{\eta}^{1\rightarrow 2}$ (resp. $\boldsymbol{\eta}^{2\rightarrow 1}$) is obtained by replacing $\mathbf{y}$ in (\ref{eq:OFDM_demod}) (resp. (\ref{eq:FBMCmod})) with the expression of $\mathbf{x}_n[k]$ in (\ref{eq:FBMC_sig}) (resp. (\ref{eq:OFDM_sig})). Then, values of $I^{1\rightarrow 2}$ (resp. $I^{2\rightarrow 1}$) are finally rated by substituting the resulting expression in (\ref{eq:I12}) (resp. (\ref{eq:I21})).
After several derivation steps, we get, as in the \textit{Hom} scenario,
\begin{eqnarray}
I^{1\rightarrow 2}_{\textit{Het}} &=& \sigma_{\mathbf{d}_1^2} \sum_{m \in \mathcal{L}_2, q \in \mathcal{L}_1} I^{1\rightarrow 2}_{\textit{Het}}(q-m),\label{eq:I12het}\\
I^{2\rightarrow 1}_{\textit{Het}} &=& \sigma_{\mathbf{d}_2^2} \sum_{m \in \mathcal{L}_1, q \in \mathcal{L}_2} I^{2\rightarrow 1}_{\textit{Het}}(q-m).\label{eq:I21het}
\end{eqnarray}
Getting mathematical closed-forms of $I^{1\rightarrow 2}_{\textit{Het}}(l)$ and $I^{2\rightarrow 1}_{\textit{Het}}(l)$ is challenging, which is why, for sake of simplicity, most studies of mutual interference in heterogeneous scenarios are based on the PSD-based model \cite{weiss04, Xu2008, shaat2010computationally,  skrzypczak2012ofdm}.

\subsection{PSD-based Interference Modeling}

\begin{figure}
	\vspace{-10pt}
	\centering
	\subfloat{\label{fig:PSD_OFDM}	\includegraphics[width=\linewidth]{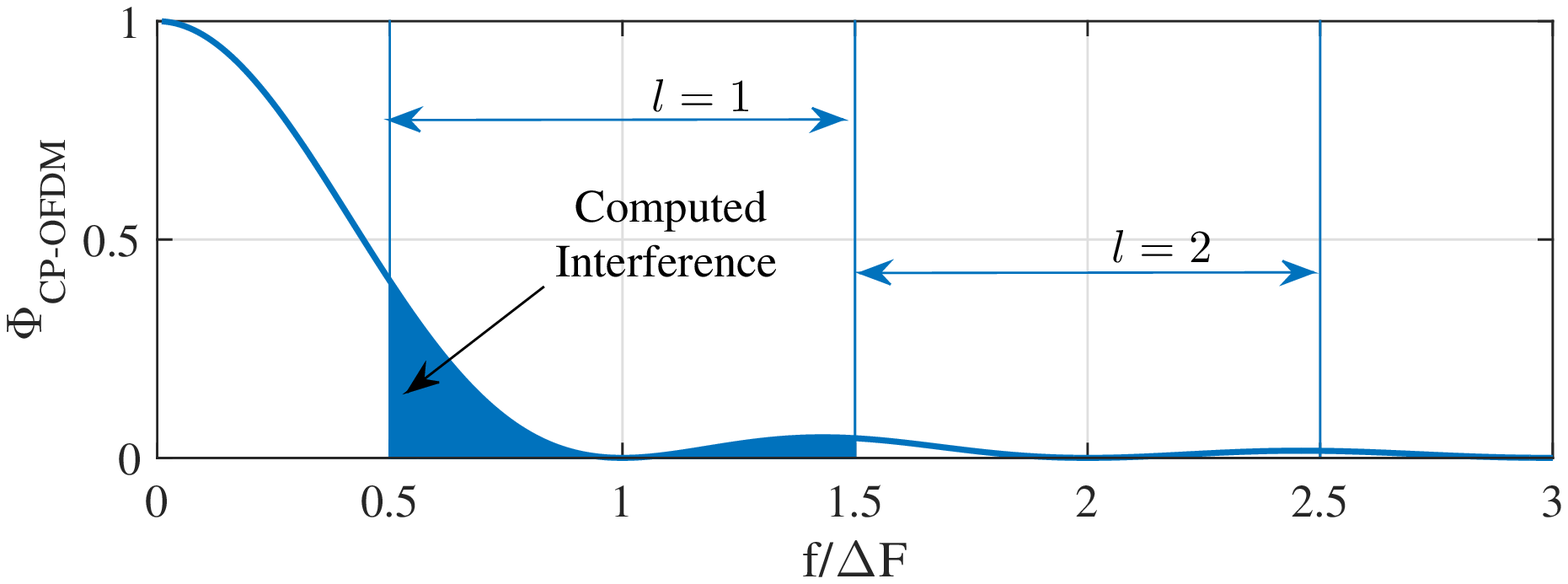}}
	\hspace{5pt}
	\subfloat{\label{fig:PSD_OQAM}	\includegraphics[width=\linewidth]{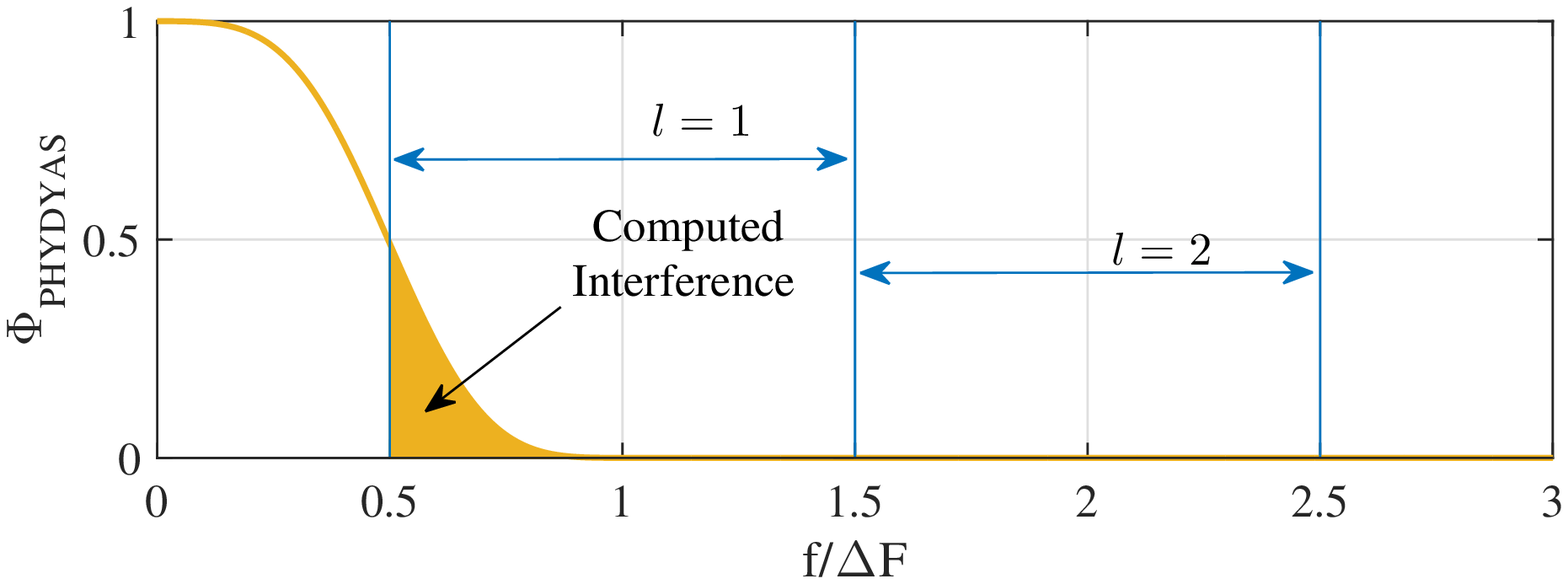}}
	\caption{Modeling of injected interference with the PSD-based model for CP-OFDM (top) and OFDM/OQAM with PHYDYAS filter (bottom). The values of interference injected by subcarrier $0$ on a subcarrier at a spectral distance of $l=1$ correspond to the integration of the PSD from $0.5\ \Delta F$ to $1.5\ \Delta F$.}
	\label{fig:PSD_model}
	\vspace{-10pt}
\end{figure}

The PSD-based model consists in computing the leakage caused by users onto each other by integrating the PSD of the interfering signal on the band that suffers from the interference. Therefore, in the \textit{Het} scenario, according to the PSD-based model, (\ref{eq:I12het}) and (\ref{eq:I21het}) are obtained by computing
\begin{eqnarray}
I_\textit{Het}^{1\rightarrow 2}(l) &=& \int\limits_{l-1/2\Delta F}^{l+1/2\Delta F}\Phi_\text{CP-OFDM}(f) df,\label{eq:PSD_mod1}\\
I_\textit{Het}^{2\rightarrow 1}(l) &=& \int\limits_{l-1/2\Delta F}^{l+1/2\Delta F}\Phi_\text{PHYDIAS}(f) df,\label{eq:PSD_mod}
\end{eqnarray}
where $\Phi_\text{CP-OFDM}$ (resp. $\Phi_\text{PHYDIAS}$ ) is the PSD of the CP-OFDM signal (resp. of OFDM/OQAM with PHYDIAS filter).
A graphical view of (\ref{eq:PSD_mod1}) and (\ref{eq:PSD_mod}) is presented in Fig.~\ref{fig:PSD_model}. It can be seen than values of interference rated with the PSD-based model are not symmetrical. As a matter of fact, because the side-lobes of $\Phi_\text{CP-OFDM}$ are much higher than those of $\Phi_\text{PHYDIAS}$, the PSD-based model will give $I_\textit{Het}^{1\rightarrow 2} \gg I_\textit{Het}^{2\rightarrow 1}$. Therefore, according to the PSD-based model, the CP-OFDM incumbent $\mathcal{U}_1$ interferes more onto the OFDM/OQAM secondary $\mathcal{U}_2$ than the opposite.

Besides, because the PSD-based model only rates the power of injected interference, it is challenging to map the obtained values of interference to higher level metrics, e.g. Bit Error Rate (BER). The only possibility offered by the PSD-based model is to approximate the statistics of heterogeneous interference as a white Gaussian noise the variance of which is given by (\ref{eq:PSD_mod}), i.e 
\begin{equation}
\boldsymbol{\eta}_m^{2\rightarrow 1} \sim \mathcal{N}(0,\sum_{q\in \mathcal{L_1}}I_\textit{Het}^{2\rightarrow 1}(q-m)).
\end{equation}
Then, classical expressions of transmission performance under white Gaussian noise in \cite{ProakisAWGN} can be applied.

\subsection{Discussing the suitability of the PSD-based model}
\label{sec:discuss}
\begin{figure}
		\vspace{-20pt}
	\includegraphics[width=\linewidth]{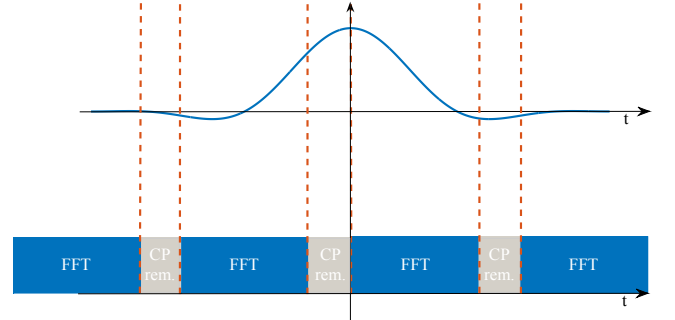}
\vspace{-60pt}
	\caption{Demodulation operations at the CP-OFDM receiver with an interfering OFDM/OQAM signal. The well-shaped PHYDYAS filter is cut in small non-contiguous parts on which FFT operations are performed.}
	\vspace{-10pt}
	\label{fig:explanation}
\end{figure}

The main pitfall of the PSD-based model lies in the fact that it does not take into account the time window of the receiver. However, this is of paramount importance as the incumbent only considers a time window with a specific width based on its own parameters. To illustrate this, Fig.~\ref{fig:explanation} shows the demodulation operations that are performed by the CP-OFDM incumbent with an interfering secondary OFDM/OQAM signal. Though the PHYDYAS filter is well spectrally localized, it has a length of  $L_\text{PHYDYAS} = KM$ samples. However, the CP-OFDM receiver window is of length $L_\text{CP-OFDM} = M$ samples. Therefore, as plotted in Fig.~\ref{fig:explanation}, the CP-OFDM incumbent demodulator performs Fast Fourier Transform (FFT) on a time window which is much shorter than the length of the prototype filter of OFDM/OQAM. In turn, the signal suffers from discontinuities that produce projections on the whole incumbent spectrum. 

Moreover, Fig.~\ref{fig:PSD_model} shows that the PSD based model consists in multiplying the interfering signal by a rectangular window in the frequency domain. In the time domain, this corresponds to filtering the interfering signal through an infinite sinc filter. Therefore, the PSD-based model does not reflect the actual demodulation operations that are processed at the CP-OFDM receiver that suffers from interference. 

Besides, Fig.~\ref{fig:explanation} shows that the prototype filter of OFDM/OQAM spans multiple time windows of the CP-OFDM incumbent receiver. Then, one OFDM/OQAM symbol interferes on several subsequent CP-OFDM symbols. This shows that the elements of $\boldsymbol{\eta}^{2\rightarrow 1}$ cannot be considered independent. Therefore, though Gaussian, the heterogeneous interference between the two users in \textit{Het} scenario is not white, but colored.

	\section{Numerical Results}
	\label{sec:results}
	In this section, we present several numerical results comparing values obtained with the PSD-based model and by numerical simulations. Besides, both scenarios \textit{Het} and \textit{Hom} are studied to rate the advantages of using OFDM/OQAM in the secondary system.

\subsection{System Setup}
We consider an incumbent system $\mathcal{U}_1$ with 3GPP LTE standard parameters with $M_a=36$ active subcarriers, which corresponds to $3$ LTE resource blocks along the frequency axis, $M = 256$ samples per symbol and $N_\text{CP} = 18$ CP samples. The secondary user $\mathcal{U}_2$ also uses $3$ LTE resource blocks along the frequency axis. No guard band is considered between the two users, and they are directly adjacent in the spectrum. More specifically, the sets of subcarriers occupied by the two users are defined as $\mathcal{L}_1 = [37\ldots 72]$ and $\mathcal{L}_2 = [73\ldots 108]$. Both users use the same subcarrier spacing $\Delta F = 15$ kHz. In \textit{Hom} scenario, CP-OFDM based $\mathcal{U}_2$ uses the same parameters as $\mathcal{U}_1$. In the \textit{Het} scenario, OFDM/OQAM based $\mathcal{U}_2$ uses $M = 256$ samples per symbol and the PHYDIAS filter with overlapping factor $K = 4$. In \textit{Het} scenario, the performance of users is evaluated though empirical estimation of (\ref{eq:I12}) and (\ref{eq:I21}) based on Monte-Carlo simulation and compared with the values expected with the PSD-based model. CP-OFDM systems transmit complex symbols drawn from a 64-Quadrature Amplitude Modulation (QAM) constellation. To ensure fairness, OFDM/OQAM systems transmit twice as much real symbols drawn from a 8-Pulse Amplitude Modulation (PAM), which corresponds to a 64 QAM after reconstruction of a complex constellation. Moreover, $\mathcal{U}_2$ starts transmitting with a delay $\tau \in [-\frac{M+N_\text{CP}}{2} \ \frac{M+N_\text{CP}}{2}[ $. Finally, numerical simulations are led on $10^5$ symbols, each carrying 6 bits. Therefore, the BER curves drawn from numerical simulation are based on a transmission of $6\times M_a \times 10^5 = 1.92\times 10^7$ bits.
 
\subsection{Interference Analysis}
\begin{figure}
	\includegraphics[width=\linewidth]{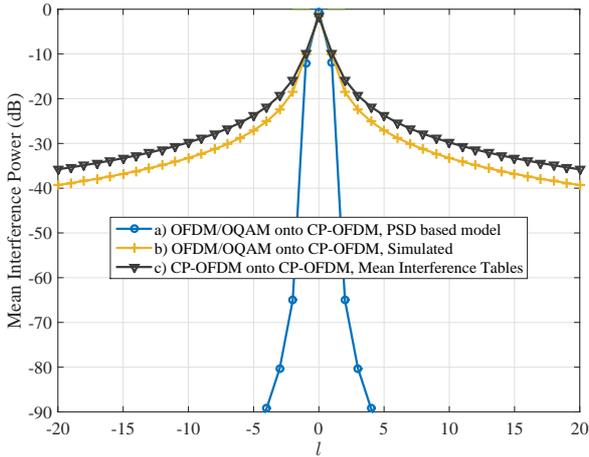}
	\caption{Comparison between interference values obtained with (a) OFDM/OQAM onto CP-OFDM with PSD-based model, (b) OFDM/OQAM onto CP-OFDM through numerical simulation, and (c) CP-OFDM onto CP-OFDM with mean interference tables \cite{Medjahdi2010a}.}
	\label{fig:interfTables}
	\vspace{-15pt}
\end{figure}

\begin{figure}
	\centering
	\subfloat[Probability Distribution Function]{\label{fig:pdf}\includegraphics[width=\linewidth]{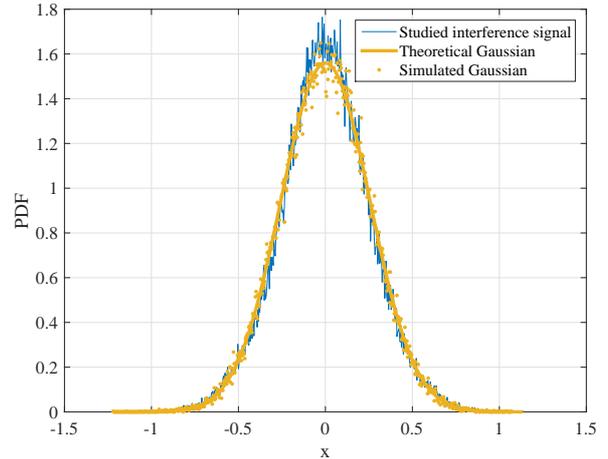}}
	\hspace{5pt}
	\subfloat[Covariance matrix]{\label{fig:cov}\includegraphics[width=\linewidth]{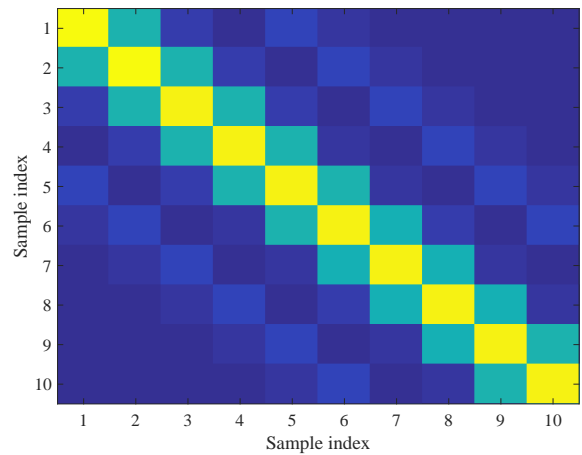}}
	\hspace{5pt}
	\caption{Statistics of interference signal caused by OFDM/OQAM onto CP-OFDM subcarrier 72. The different results show that it can be modeled by a colored Gaussian noise.}
	\label{fig:stats}
	\vspace{-15pt}
\end{figure}

First, we aim to rate the interference caused by $\mathcal{U}_2$ on the incumbent $\mathcal{U}_1$. Fig.~\ref{fig:interfTables} presents the values of $I^{2\rightarrow 1}(l)$ in dB for spectral distance $l \in [-20 \ 20]$. For the \textit{Het} scenario, i.e. when $\mathcal{U}_2$ uses OFDM/OQAM, we present values obtained with both the PSD-based model and through numerical simulations. We can observe in Fig.~\ref{fig:interfTables} a tremendous gap between the values of interference planned by the PSD-based model and the real ones. As a case in point, at $l = 2$, the PSD-based model plans that the value of the interference injected on the incumbent will be about $-65$ dB, whereas numerical simulations show that the actual interference value is $-18.5$ dB. Moreover, for $l = 20$, the PSD-based model predicts that the injected interference will be insignificant, whereas the numerical simulations show that it is still at a non-negligible level of $-40$ dB. This proves that in the \textit{Het} scenario, the PSD-based model completely fails to give a good approximation of the interference injected by an OFDM/OQAM secondary user onto an incumbent CP-OFDM system.

Besides, the mean interference tables from \cite{Medjahdi2010a} are plotted to rate the interference injected by $\mathcal{U}_2$ onto $\mathcal{U}_1$ for the \textit{Hom} scenario, when both systems are using CP-OFDM. It shows that the interference injected onto $\mathcal{U}_1$ can be reduced by approximately $5$ dB if $\mathcal{U}_2$ uses OFDM/OQAM. Though much less than what was expected with the PSD-based model, this gain is still high enough to be noticed.

Having rated the power of injected interference, we focus now on the statistics of the latter in the \textit{Het} scenario. To do so, we scrutinize the distribution of $\boldsymbol{\eta}_{72}^{2\rightarrow 1}$, which corresponds to the interference injected by the OFDM/OQAM based $\mathcal{U}_2$ onto the closest subcarrier of $\mathcal{U}_1$. We show the Probability Distribution Function (PDF) of this interference signal in Fig.~\ref{fig:pdf}. We can notice that it is well approximated by a Gaussian function of variance $I_{72,\textit{Het}}^{2\rightarrow 1}$. However, the covariance matrix of the studied interference, plotted in Fig.~\ref{fig:cov} is band-diagonal. This reveals a dependency between subsequent samples of the interference signal $\boldsymbol{\eta}_{72}^{2\rightarrow 1}$. These two figures therefore corroborate the remarks we highlighted in Section \ref{sec:discuss} and confirm the fact that heterogeneous interference is colored.

\subsection{Transmission Performance}

\begin{figure}
	\vspace{-10pt}
	\includegraphics[width=\linewidth]{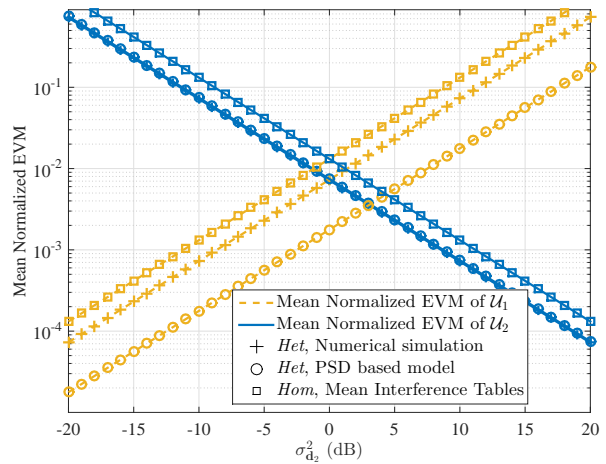}
	\caption{Mean Normalized EVM of users $\mathcal{U}_1$ and $\mathcal{U}_2$ in the scenarios \textit{Het} and \textit{Hom}}
	\label{fig:EVM}
\end{figure}

We now focus on the transmission performance of both users. To do so, we set the power of the symbols transmitted by the incumbent system $\mathcal{U}_1$ as $\sigma_{\mathbf{d}_1}^2 = 0$ dB and we sweep $\sigma_{\mathbf{d}_2}^2$ from $-20$ dB to $20$ dB. Here, we focus on the effects of inter-user interference caused by  the adjacent transmissions of the two users. Therefore, no channel and no noise is considered.
The normalized Error Vector Magnitude (EVM) obtained for both users is plotted in Fig.~\ref{fig:EVM}. Here, our observations are threefold: first, the PSD-based model approximates surprisingly well the interference seen by the secondary OFDM/OQAM user $\mathcal{U}_2$ in the \textit{Het} scenario. This shows that the PSD-based model may still be suitable in some cases, especially when the time window of the receiver is longer than the interfering signal. However, the PSD-based model dramatically underestimates the interference seen by the incumbent receiver $\mathcal{U}_1$. Second, we point out that the actual inter-user interference in the \textit{Het} scenario is symmetrical. As a case in point, the obtained EVM values for both users are equal when their transmission power is equal. This contradicts the PSD-based model, which predicts that the incumbent CP-OFDM $\mathcal{U}_1$ will be more protected than the secondary OFDM/OQAM $\mathcal{U}_2$ :  according to the PSD-based model, the normalized EVM values of both users are equal when $\sigma_{\mathbf{d}_2^2} = 3$ dB. Third, both $\mathcal{U}_1$ and $\mathcal{U}_2$ experience lower EVM in the \textit{Het} scenario than in the \textit{Hom} scenario.

\begin{figure}
	\vspace{-10pt}
	\includegraphics[width=\linewidth]{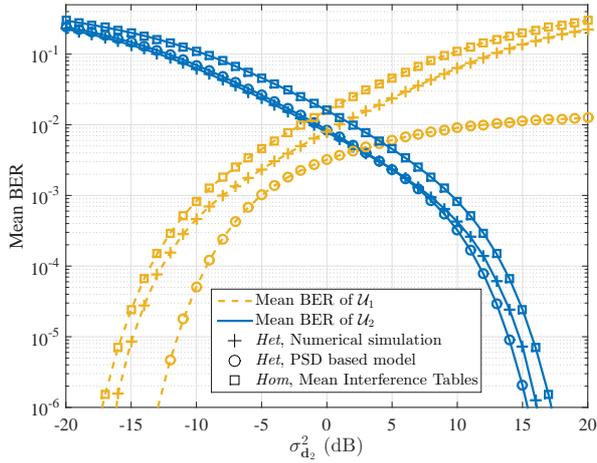}
	\caption{Mean BER of users $\mathcal{U}_1$ and $\mathcal{U}_2$ in the scenarios \textit{Het} and \textit{Hom}}
	\label{fig:BER}
\end{figure}

Based on the above results, we analyze the BER for both users in Fig.~\ref{fig:BER}. As said in Section  \ref{sec:discuss}, the interfering signal is approximated to a white Gaussian noise to compute the BER from the EVM for both the PSD-based model in the \textit{Het} scenario and the instantaneous interference tables in the \textit{Hom} scenario. This allows to compute BER thanks to the classical expressions of the BER of M-ary QAM constellations \cite{ProakisAWGN}.
As expected, the obtained BER performance confirms the EVM behaviour presented in Fig.\ref{fig:EVM}, and again, 
the PSD-based model is totally wrong for modeling the interference seen by the incumbent $\mathcal{U}_1$ in the \textit{Het} scenario. Nevertheless, it gives a satisfying approximation of the BER of OFDM/OQAM based $\mathcal{U}_2$, especially for values of BER higher than $10^{-3}$. However, when the BER of OFDM/OQAM based $\mathcal{U}_2$ becomes low (for $\sigma_{\mathbf{d}_2}^2 > 10 $ dB), the PSD-based model understimates it. This is due to the fact that the interference was approximated as a white gaussian signal whereas it has been shown that it is colored in Fig.~\ref{fig:cov}. Finally, Fig.~\ref{fig:BER} shows that the benefits of using OFDM/OQAM are not as high as what was expected with the PSD-based model. Yet, it shows that using OFDM/OQAM for the secondary $\mathcal{U}_2$ does still bring some advantage. For example, when both users have the same transmission power, the BER of each user in the \textit{Het} scenario is equal to half what they experience
in the \textit{Hom} scenario.

\begin{figure}
	\vspace{-10pt}
	\includegraphics[width=\linewidth]{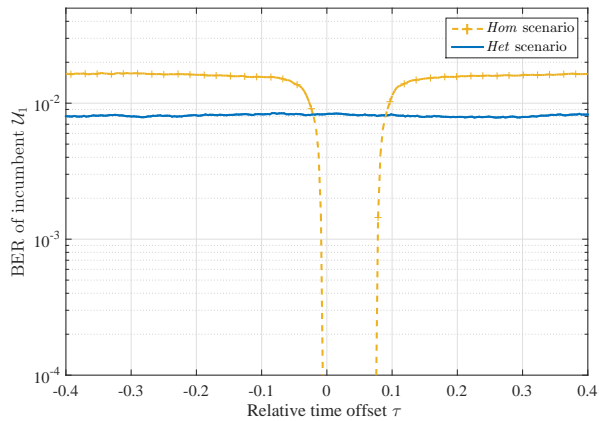}
	\caption{Empirical BER of $\mathcal{U}_1$ versus the timing offset $\tau$ between $\mathcal{U}_2$ and $\mathcal{U}_1$, for \textit{Het} and \textit{Hom} scenarios.}
	\label{fig:BER_TO}
\end{figure}

Presented results show that, in scenarios where an incumbent CP-OFDM system coexists with an asynchronous user $\mathcal{U}_2$, it is still advantageous to both users that $\mathcal{U}_2$ uses OFDM/OQAM, though benefits are much less important than those planned with the PSD-based model in \cite{shaat2010computationally, skrzypczak2012ofdm}.
To conclude this study, we focus on the BER of the incumbent $\mathcal{U}_1$ in both \textit{Het} and \textit{Hom} scenarios with a deterministic value of $\tau$. 
Fig.\ref{fig:BER_TO} highlights three different and interesting results: first, in the \textit{Het} scenario, $\tau$ has no impact. This is mainly due to the fact that OFDM/OQAM and CP-OFDM systems are inherently asynchronous, as they do not have the same time spacing between subsequent symbols (see Fig.\ref{fig:explanation}). Second, in the \textit{Hom} scenario, if the timing offset can be contained in the CP duration, the performance of incumbent $\mathcal{U}_1$ is not degraded at all. This is a well known result concerning Multi-User Interference in CP-OFDM systems. Third, as soon as $\tau$ grows higher than the CP duration, it is worth using OFDM/OQAM instead of CP-OFDM at the secondary system $\mathcal{U}_2$ to protect $\mathcal{U}_1$.

	\section{Conclusion}
	\label{sec:ccl}
	In this paper, we analyzed a scenario in which the coexistence between a legacy CP-OFDM incumbent system and an asynchronous secondary user produces inter-user interference. We analyzed the performance of users in the cases where the secondary user utilizes CP-OFDM or OFDM/OQAM waveform.

We showed that the widely used PSD-based model is highly flawed and fails to give a good approximation of the interference seen by each user in heterogeneous scenarios. Indeed, presented numerical results showed that when the secondary system utilizes OFDM/OQAM, the actual values of interference are higher than those planned by the PSD based model by more than $50$ dB. Furthermore, contrary to the widely spread idea that CP-OFDM interferes more onto OFDM/OQAM users than the opposite, we revealed that heterogeneous interference is symmetrical and that users interfere equally onto each other.

Though it was shown that both users experience a slight improvement when the secondary user uses OFDM/OQAM modulation, the gain was shown to be much more limited than what was expected with the PSD-based model.

To conclude, we showed in this paper that models existing in the literature to rate interference in heterogeneous networks are not satisfying. Future work will therefore focus on deriving analytical closed-forms of heterogeneous interference that can be used to extensively investigate such scenarios.

	\normalsize
	\bibliographystyle{IEEEtran}
	\bibliography{IEEEabrv,main}

\begin{thebibliography}{10}
\providecommand{\url}[1]{#1}
\csname url@samestyle\endcsname
\providecommand{\newblock}{\relax}
\providecommand{\bibinfo}[2]{#2}
\providecommand{\BIBentrySTDinterwordspacing}{\spaceskip=0pt\relax}
\providecommand{\BIBentryALTinterwordstretchfactor}{4}
\providecommand{\BIBentryALTinterwordspacing}{\spaceskip=\fontdimen2\font plus
\BIBentryALTinterwordstretchfactor\fontdimen3\font minus
  \fontdimen4\font\relax}
\providecommand{\BIBforeignlanguage}[2]{{%
\expandafter\ifx\csname l@#1\endcsname\relax
\typeout{** WARNING: IEEEtran.bst: No hyphenation pattern has been}%
\typeout{** loaded for the language `#1'. Using the pattern for}%
\typeout{** the default language instead.}%
\else
\language=\csname l@#1\endcsname
\fi
#2}}
\providecommand{\BIBdecl}{\relax}
\BIBdecl

\bibitem{Lahetkangas2013}
E.~L\"{a}hetkangas and H.~Lin, ``{Deliverable D2.1 - Requirement analysis and
  design approaches for 5G air interface},'' p.~72, 2013.

\bibitem{wunder2014}
G.~Wunder, P.~Jung, M.~Kasparick, T.~Wild, F.~Schaich, Y.~Chen, S.~Brink,
  I.~Gaspar, N.~Michailow, A.~Festag, L.~Mendes, N.~Cassiau, D.~Ktenas,
  M.~Dryjanski, S.~Pietrzyk, B.~Eged, P.~Vago, and F.~Wiedmann, ``{5GNOW:
  non-orthogonal, asynchronous waveforms for future mobile applications},''
  \emph{IEEE Communications Magazine}, vol.~52, no. February, pp. 97--105,
  2014.

\bibitem{Medjahdi2011}
Y.~Medjahdi, M.~Terr\'{e}, D.~L. Ruyet, D.~Roviras, and A.~Dziri,
  ``{Performance analysis in the downlink of asynchronous OFDM/FBMC based
  multi-cellular networks},'' \emph{IEEE Transactions on Wireless
  Communications}, vol.~10, no.~8, pp. 2630--2639, 2011.

\bibitem{Farhang-Boroujeny2011}
B.~Farhang-Boroujeny, ``{OFDM versus filter bank multicarrier},'' \emph{Signal
  Processing Magazine, IEEE}, no. MAY 2011, pp. 92--112, 2011.

\bibitem{Baltar2007}
L.~G. Baltar, D.~S. Waldhauser, and J.~A. Nossek, ``{Out-Of-Band Radiation in
  Multicarrier Systems: A Comparison},'' in \emph{Multi-Carrier Spread Spectrum
  2007}.\hskip 1em plus 0.5em minus 0.4em\relax Springer Netherlands, 2007,
  vol.~1, pp. 107--116.

\bibitem{Mahmoud2009}
H.~Mahmoud, T.~Yucek, and H.~Arslan, ``{OFDM for cognitive radio: merits and
  challenges},'' \emph{IEEE Wireless Communications}, vol.~16, no. April, pp.
  6--15, 2009.

\bibitem{Raghunath2009}
K.~Raghunath and a.~Chockalingam, ``{SC-FDMA versus OFDMA: Sensitivity to large
  carrier frequency and timing offsets on the uplink},'' \emph{GLOBECOM - IEEE
  Global Telecommunications Conference}, 2009.

\bibitem{Aminjavaheri2015}
A.~Aminjavaheri, A.~Farhang, A.~RezazadehReyhani, and B.~Farhang-Boroujeny,
  ``Impact of timing and frequency offsets on multicarrier waveform candidates
  for 5g,'' in \emph{Signal Processing and Signal Processing Education Workshop
  (SP/SPE), 2015 IEEE}, Aug 2015, pp. 178--183.

\bibitem{Speth1999}
M.~Speth, S.~a.~S. Fechtel, G.~Fock, H.~Meyr, and S.~Member, ``{Optimum
  receiver design for wireless broad-band systems using OFDM. I},'' \emph{IEEE
  Transactions on Communications}, vol.~47, no.~11, pp. 1668--1677, 1999.

\bibitem{Bellanger2010}
M.~Bellanger, ``{FBMC physical layer: a primer},'' pp. 1--31, 2010.

\bibitem{Xing2014}
H.~Xing and M.~Renfors, ``{Investigation of filter bank based device-to-device
  communication integrated into OFDMA cellular system},'' in \emph{ISWCS},
  2014, pp. 513--518.

\bibitem{BodinierICC2016}
Q.~Bodinier, A.~Farhang, F.~Bader, H.~Ahmadi, J.~Palicot, and L.~A. DaSilva,
  ``{5G Waveforms for Overlay D2D Communications: Effects of Time-Frequency
  Misalignment},'' in \emph{IEEE International Conference on Communications
  (ICC)}, Kuala Lumpur, 2016 (Accepted).

\bibitem{Xu2008}
R.~Xu and M.~Chen, ``{Spectral leakage suppression of DFT based OFDM via
  adjacent subcarriers correlative coding},'' in \emph{Proccedings of IEEE
  Global Telecommunications Conference (GLOBECOM '08) December 2008.}, 2018,
  pp. 3029--3033.

\bibitem{shaat2010computationally}
M.~Shaat and F.~Bader, ``Computationally efficient power allocation algorithm
  in multicarrier-based cognitive radio networks: Ofdm and fbmc systems,''
  \emph{EURASIP Journal on Advances in Signal Processing}, vol. 2010, p.~5,
  2010.

\bibitem{skrzypczak2012ofdm}
A.~Skrzypczak, J.~Palicot, and P.~Siohan, ``{OFDM/OQAM modulation for efficient
  dynamic spectrum access},'' \emph{International Journal of Communication
  Networks and Distributed Systems}, vol.~8, no. 3-4, pp. 247--266, 2012.

\bibitem{weiss04}
A.~K. T.~Weiss, J.~Hillenbrand and F.~K. Jondral, ``{Mutual Interference in
  OFDM-Basedased Spectrum Pooling Systems,},'' in \emph{Proceedings of the 59th
  IEEE Vehicular Technology Conference (VTC '04). Milan, Italy.}, vol.~59,
  2004.

\bibitem{Medjahdi2010a}
Y.~Medjahdi, M.~Terr\'{e}, D.~L. Ruyet, and D.~Roviras, ``{Interference tables:
  a useful model for interference analysis in asynchronous multicarrier
  transmission},'' \emph{EURASIP Journal on Advances in Signal Processing},
  vol. 2014, no.~54, pp. 1--17, 2014.

\bibitem{Bellanger}
M.~Bellanger, ``{Specification and design of a prototype filter for filter bank
  based multicarrier transmission},'' in \emph{2001 IEEE International
  Conference on Acoustics, Speech, and Signal Processing. Proceedings}, vol.~4,
  2001, pp. 2417--2420.

\bibitem{ProakisAWGN}
J.~G. Proakis, ``{Optimum Receivers for the Additive White Gaussian Noise
  Channel},'' in \emph{Digital Communications}, internatio~ed., McGraw-Hill,
  Ed., 2001, pp. 231--332.

\end{thebibliography}

\end{document}